\documentclass[a4paper]{jpconf}
\usepackage{graphicx}
\begin{document}
\title{Constraints on small-scale cosmological perturbations from gamma-ray searches for dark matter}
\author{Pat Scott$^1$, Torsten Bringmann$^2$ \& Yashar Akrami$^{3,4}$}
\address{$^1$Department of Physics, McGill University, 
3600 rue University, Montr\'eal, QC, H3A 2T8, Canada, \texttt{patscott@physics.mcgill.ca}}
\address{$^2$II.\ Institute for Theoretical Physics, University of Hamburg, Luruper Chausse 149, DE-22761 Hamburg, Germany, \texttt{torsten.bringmann@desy.de}}
\address{$^3$Institute of Theoretical Astrophysics, University of Oslo, P.O. Box 1029 Blindern, N-0315 Oslo, Norway, \texttt{yashar.akrami@astro.uio.no}}
\address{$^4$The Oskar Klein Centre for Cosmoparticle Physics, Department of Physics, Stockholm University, AlbaNova, SE-106 91 Stockholm, Sweden}

\begin{abstract}
Events like inflation or phase transitions can produce large density perturbations on very small scales in the early Universe.  Probes of small scales are therefore useful for e.g. discriminating between inflationary models.  Until recently, the only such constraint came from non-observation of primordial black holes (PBHs), associated with the largest perturbations.  Moderate-amplitude perturbations can collapse shortly after matter-radiation equality to form ultracompact minihalos (UCMHs) of dark matter, in far greater abundance than PBHs.  If dark matter self-annihilates, UCMHs become excellent targets for indirect detection.  Here we discuss the gamma-ray fluxes expected from UCMHs, the prospects of observing them with gamma-ray telescopes, and limits upon the primordial power spectrum derived from their non-observation by the \textit{Fermi} Large Area Space Telescope.
\end{abstract}

\begin{figure}[t]
\includegraphics[width=0.5\columnwidth, trim = 0 20 0 0, clip=true]{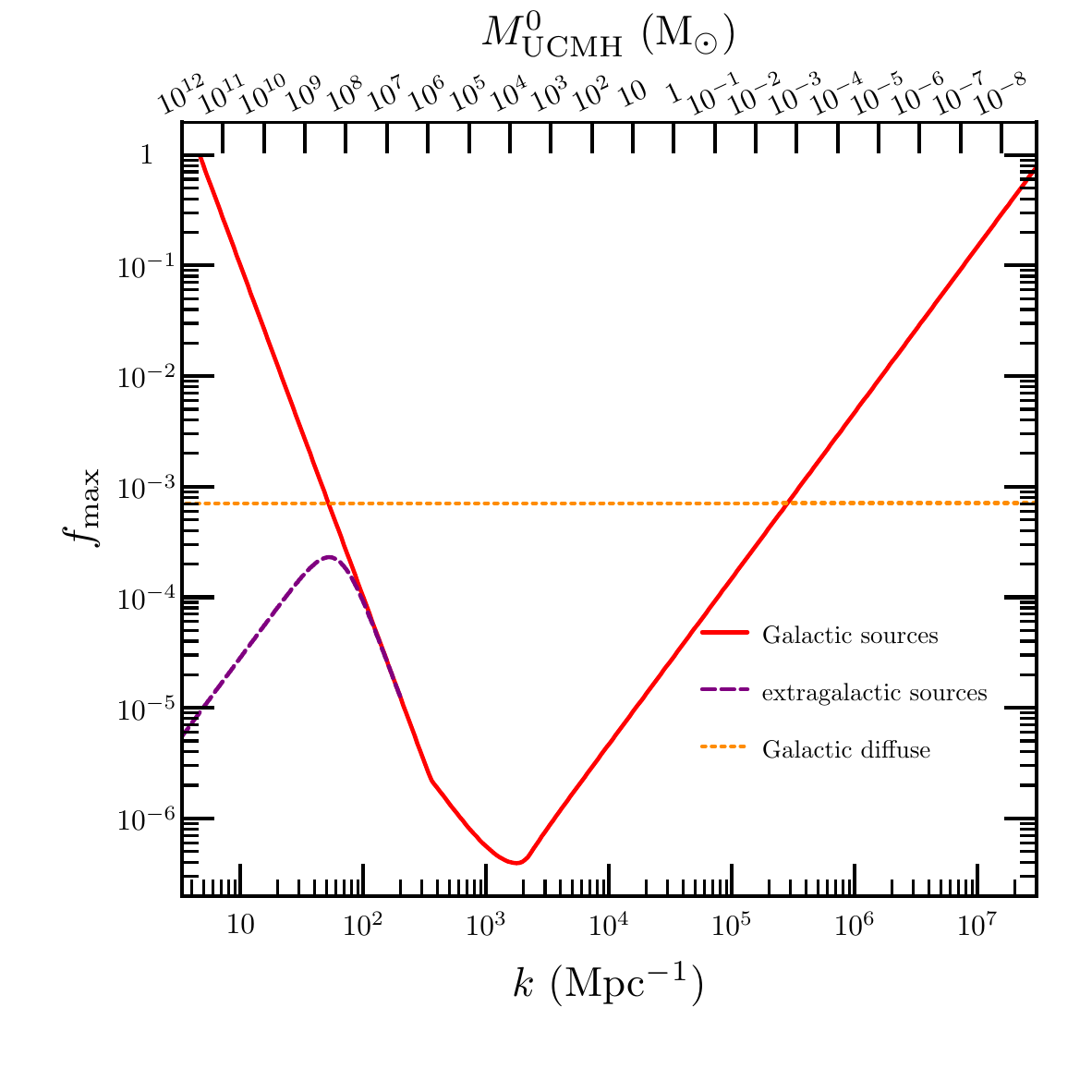}
\includegraphics[width=0.5\columnwidth, trim = 0 20 0 0, clip=true]{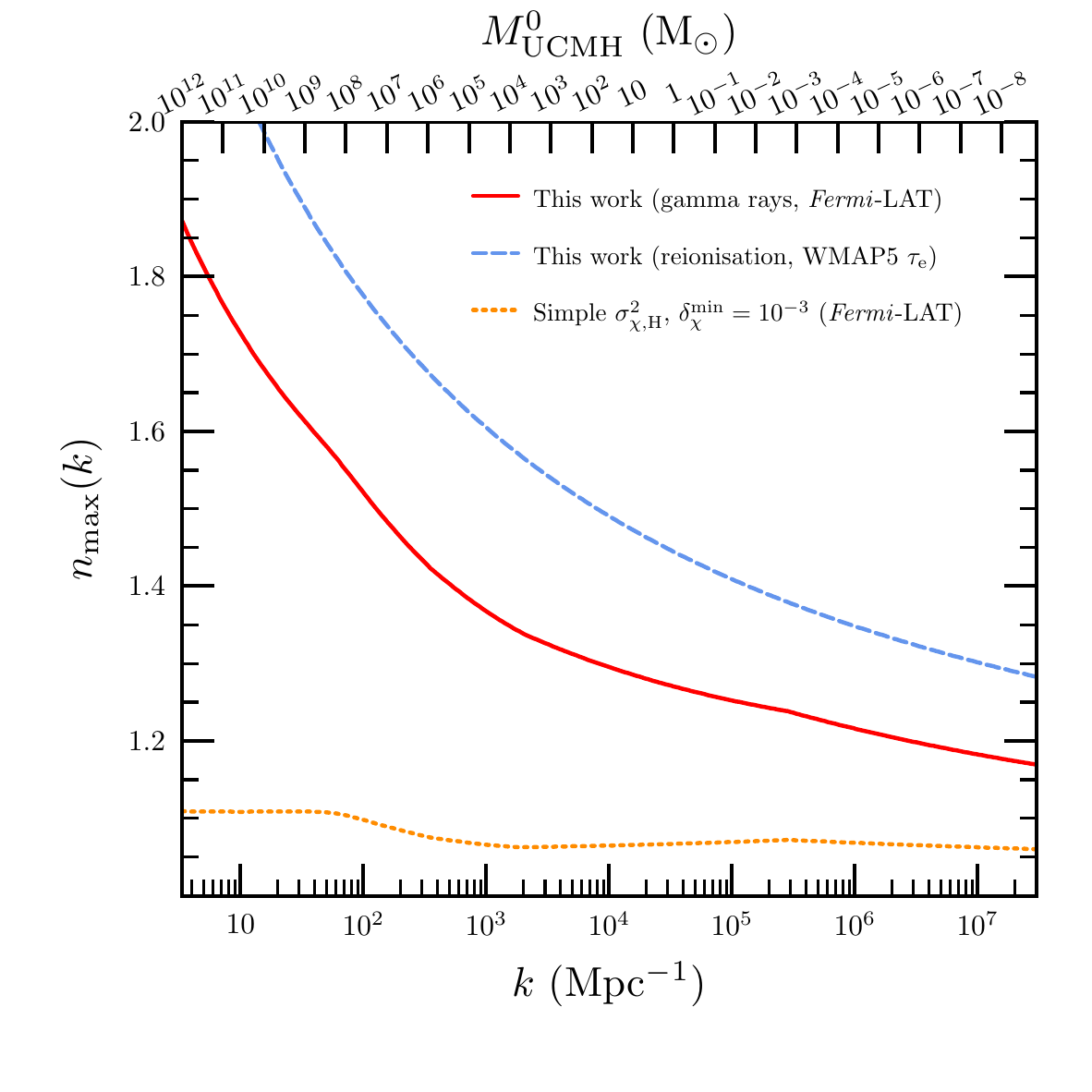}
\caption{\textit{Left}: Limits on the abundance of UCMHs in the region around our Galaxy, given as a fraction of DM in UCMHs.  We show 95\% CL upper limits from \textit{Fermi}-LAT searches for individual and diffuse DM sources.  The limits arising from searches for individual minihalos are based on one year of \textit{Fermi} operation in all-sky survey mode.  \textit{Right}: The largest allowed spectral index of the primordial power spectrum, as determined by gamma-ray searches for UCMHs, and the impact of UCMHs on reionisation.  Here we assume $\delta^2_\mathrm{H}\propto k^{n-1}$, and take into account the best limits on $f_\mathrm{UCMH}$ given in the left panel, for wave numbers smaller than $k$.  We also give the resulting gamma-ray constraint if we were to assume $\delta_\chi^\mathrm{min}=10^{-3}$ (improved upon in this work; see \cite{Bringmann11UCMH}), and a simplified estimate of $\sigma_{\chi,\mathrm{H}}$ \cite{GreenLiddle} (corrected in this work; see \cite{Bringmann11UCMH}).  From \cite{Bringmann11UCMH}.}
\label{fig:1}
\end{figure}

\begin{figure}[t]
\includegraphics[width=0.5\columnwidth, trim = 0 20 0 0, clip=true]{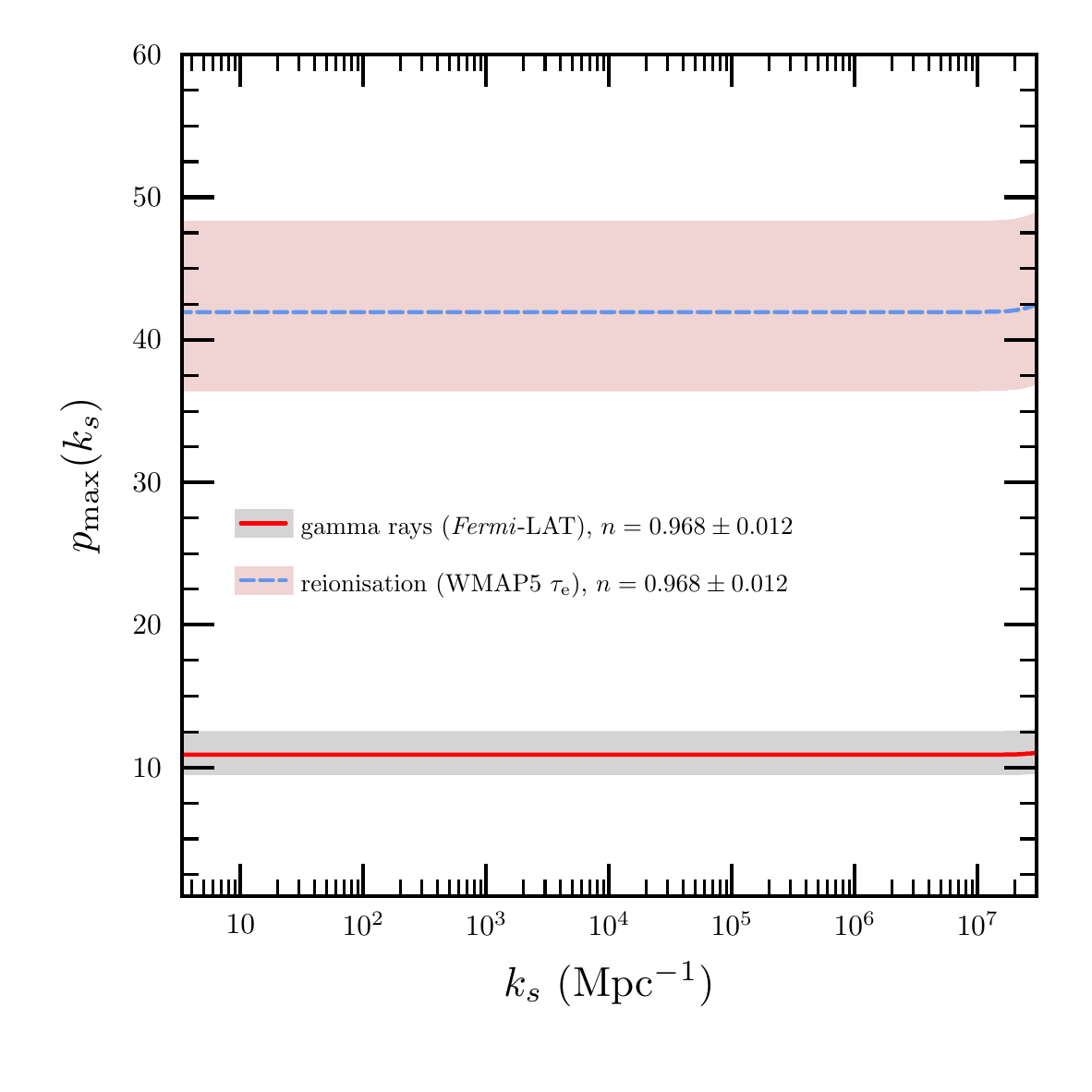}
\includegraphics[width=0.5\columnwidth, trim = 0 20 0 0, clip=true]{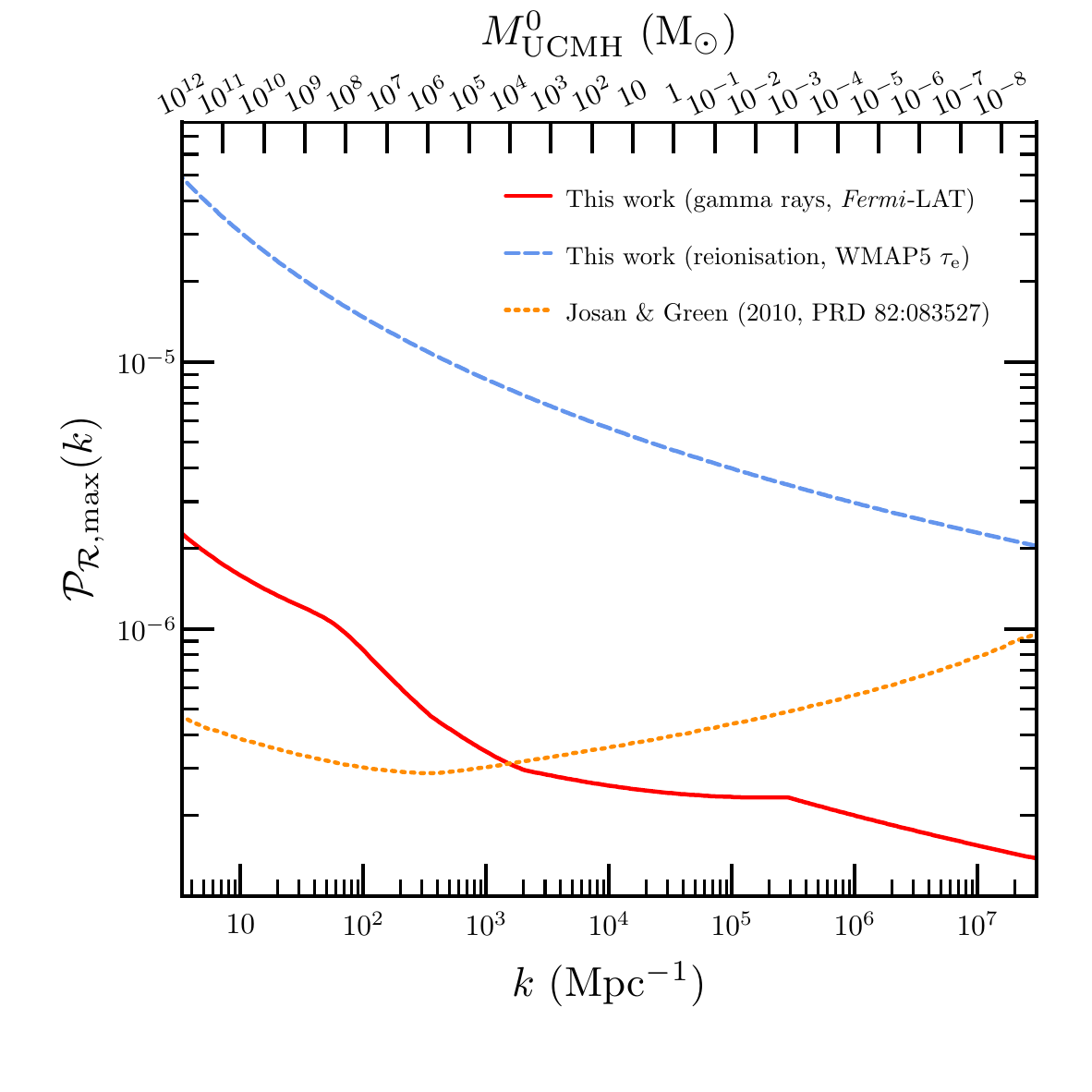}
\caption{\textit{Left}: The allowed height $p$ of a step in the primordial power spectrum, as determined by gamma-ray searches for UCMHs and impacts on reionisation, as a function of the location $k_s$ of the step.  The parameter $p$ refers to the ratio of power at wavenumbers immediately above and below $k_s$.  Main curves assume $n=0.968$ (obtained from WMAP7 observations of large scales \cite{wmap}), and shaded areas correspond to a 68\% CL for this measurement ($\Delta n=0.012$).  \textit{Right}: Upper limits at 95\% confidence level on the amplitude of primordial curvature perturbations $\mathcal{P}_\mathcal{R}$, from potential impacts on reionisation and non-observation of UCMHs by \textit{Fermi}.  Here we assume a non-parametric, generalised spectrum.  We also show limits from \emph{Fermi} non-observation of UCMHs derived in Ref.~\cite{JG10} for comparison, which include a more simplified treatment of the mass variance, observational statistics and minimum density contrast required to form a UCMH.  Constraints on the generalised amplitude of primordial density perturbations are found by multiplying these limits by a factor of 0.191.  From \cite{Bringmann11UCMH}.}
\label{fig:2}
\end{figure}

\begin{figure}[t]
\includegraphics[width=\linewidth, trim = 0 92 0 100, clip=true]{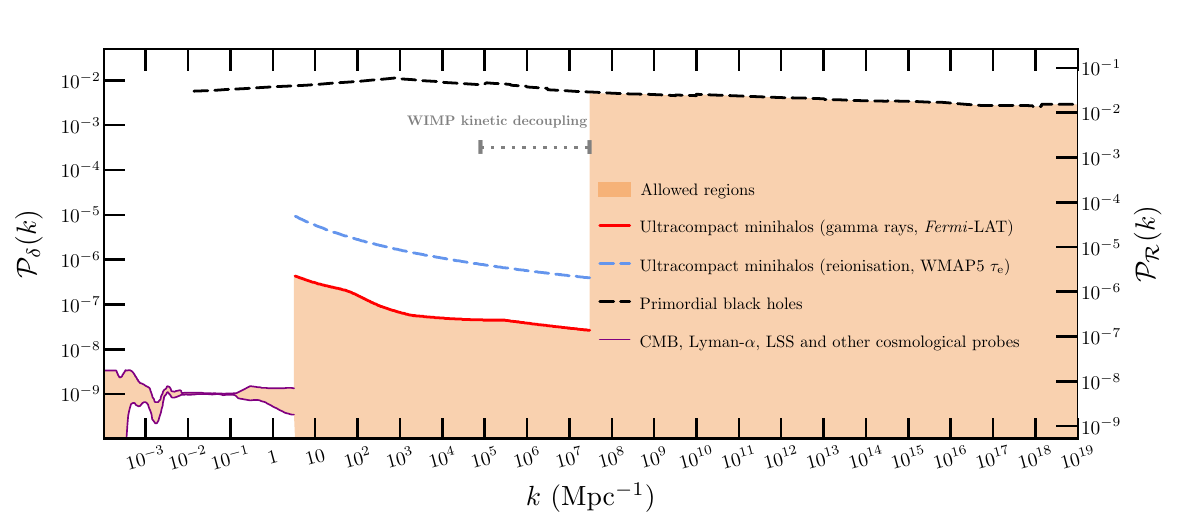}
\caption{The allowed amplitude of primordial curvature (density) perturbations $\mathcal{P}_\mathcal{R}$ ($\mathcal{P}_\delta$) at all scales.  We give the overall best measurements of the power spectrum on large scales from the CMB, large scale structure (LSS), Lyman-$\alpha$ observations and other cosmological probes \cite{Nicholson:2009pi,Nicholson:2009zj,Bird:2010mp}, as well as upper limits from searches for UCMHs with gamma-rays and the CMB.  We also show limits from searches for PBHs \cite{JGM09}. In addition, we give possible DM kinetic decoupling scales for supersymmetric WIMP models \cite{Bringmann:2009vf}.  For particle models with kinetic decoupling scales $k_\mathrm{KD}$, no UCMH limits apply for $k>k_\mathrm{KD}$.  From \cite{Bringmann11UCMH}.}
\label{fig:3}
\end{figure}

Recent work \cite{Ricotti:2009bs,Bringmann11UCMH} has shown that moderate-amplitude density perturbations produced in the early Universe can collapse to form ultra-dense halos of dark matter (DM), known as ultracompact minihalos (UCMHs).  It was quickly realised \cite{SS09,Lacki:2010zf,JG10} that such compact objects present an excellent target for indirect searches for self-annihilating DM, as they exhibit extremely steep inner radial density profiles ($\rho\propto r^{-\frac94}$).  This profile comes about because UCMHs collapse very shortly after matter-radiation equality, and therefore form essentially in isolation, by almost pure radial infall \cite{Ricotti:2009bs} (see also \cite{Bertschinger:1985pd,Vogelsberger:2009bn}).  

In a longer paper \cite{Bringmann11UCMH}, we investigated the abundance of UCMHs allowed in and around our Galaxy by gamma-ray and cosmic microwave background (CMB) observations, in order to determine implications for the allowed amplitude of cosmological perturbations in the early Universe.  We summarise those findings here; the reader is referred to the longer paper for full details of all calculations.  All calculations shown in these proceedings assume that UCMHs collapse at $z=1000$, a DM mass of 1\,TeV, and 100\% annihilation into $\bar b b$ pairs with the standard thermal cross-section ($3\times10^{-26}$\,cm$^3$\,s$^{-1}$); we note, however, that the dependence of these limits on the properties of the annihilating particles is actually rather weak.

In the left panel of Fig.~\ref{fig:1}, we show the allowed fraction of DM that might be in UCMHs, as given by three different gamma-ray observations by the \textit{Fermi} Large Area Telescope (LAT): searches for individual (point and extended) DM sources within our own Galaxy, searches for similar sources outside the Galaxy, and the total gamma-ray diffuse flux.  Here we base our limits on the non-observation of any DM sources by \textit{Fermi}, and the demand that the diffuse flux from UCMHs not exceed the total diffuse gamma-ray flux observed by \textit{Fermi} \cite{2FGL}.  At very large scales, the strongest limits come from the extragalactic search, at moderate scales from the Galactic source search, and at the smallest scales, the best limits are provided by the diffuse flux measurement.  We translate these limits into limits on the spectral index of primordial perturbations $n$ (right panel of Fig.~\ref{fig:1}), in the scenario where perturbations follow a simple unbroken power law (as might e.g. be expected from the simplest inflationary models).  The best limit on $n$ ($n\le 1.17$) is rather tight, given that UCMHs probe scales many orders of magnitude below those accessible to the CMB.  For comparison, for primordial black holes (PBHs) this limit is $n\le1.27$ \cite{astro-ph/0109404}).  In this panel (and the following ones), we also give limits derived from the impact of UCMHs on reionisation, and the resulting change of the integrated CMB optical depth due to electron scattering \cite{Zhang:2010cj}.  The reionisation limits provide support to those from gamma-rays, but are significantly weaker in general.

In the left panel of Fig.~\ref{fig:2}, we instead consider models where the spectrum follows a power law but has a step at some characteristic scale $k_\mathrm{s}$ (left panel), constraining the height $p$ of this step to be less than a factor of about 11 in the spectral amplitude.  In the right-hand panel of Fig.~\ref{fig:2}, we give limits on the raw power spectrum of curvature perturbations, using a generalised, non-parametric form for the spectrum of primordial perturbations.  The final limits are compared with earlier limits from UCMHs \cite{JG10} in this panel, and with all other limits or measurements of the primordial power spectrum in Fig.~\ref{fig:3}.  From Fig.~\ref{fig:3} it is clear that UCMHs provide an extremely valuable limit on the size of cosmological perturbations at scales between $k\sim3\,{\rm Mpc}^{-1}$ and $k\sim10^7\,{\rm Mpc}^{-1}$, well below any scale probed by the CMB.  

The new limits we present here on the primordial power spectrum are up to 5 orders of magnitude better than the previous best limits for these scales, provided by PBHs \cite{JGM09}.  Further improvements are expected to come from improved gamma-ray observations, and more sensitive microlensing searches for UCMHs (e.g.~\cite{Erickcek10,Griest11,Erickcek12}).  The latter will be especially important, as unlike gamma-ray limits, microlensing does not rely on the assumption that DM can self-annihilate.

\ack
We warmly thank Venya Berezinsky, Adrienne Erickcek, Sofia Sivertsson and Alexander Westphal for enlightening discussions. P.S. thanks the Astroparticle Group of the II.\ Institute for Theoretical Physics at the University of Hamburg for their hospitality, with support from the LEXI initiative Hamburg, and is supported by the Lorne Trottier Chair in Astrophysics and an Institute for Particle Physics Theory Fellowship. T.B.\ acknowledges support from the
German Research Foundation (DFG) through the Emmy Noether grant BR 3954/1-1. Y.A.\ thanks the Swedish Research Council (VR) for financial support and was partially supported by the European Research Council (ERC) Starting Grant StG2010-257080.

\end{document}